# Leaky mode analysis of luminescent thin films: the case of ZnO on sapphire

Roy AAD[1*], Laurent DIVAY[1+], Aurelien BRUYANT[1], Sylvain BLAIZE[1], Christophe COUTEAU[1], Dave John ROGERS[2] and Gilles LERONDEL[1*]

[1] *Laboratoire de Nanotechnologie et d'Instrumentation Optique, Institut Charles Delaunay, CNRS UMR 6279, Université de Technologie de Troyes, 12 rue Marie Curie, BP 2060, 10010 Troyes Cedex, France.*

[2] *Nanovation, 8 route de Chevreuse, Châteaufort, 78117, France.*

**Abstract**

Zinc oxide (ZnO) epitaxial thin films grown on c-sapphire substrates by pulsed laser deposition were investigated using angle and polarization-resolved photoluminescence spectroscopy. Side-emission spectra differed significantly from surface-emission spectra in exhibiting dominant, narrow, polarization-resolved peaks. These spectral features were attributed to leaky substrate modes in the layers. Observations were first verified using transmission calculations with non-adjustable parameters, which took into account the dispersion, the anisotropy of the ZnO refractive index and the dependence on film thickness. Results were consistent with Fabry-Perot-like interference being the origin of the distinctive ZnO luminescence observed at grazing incidence angles. A second analysis, based on the source terms method, was used in order to retrieve the bulk emission properties, including the wavelength-dependent quantum yield and the emission anisotropy. While ZnO thin films were considered here, this analysis method can be extended to any luminescent thin film of similar geometry, demonstrating the potential of leaky mode analysis for probing passive and active material properties.

\* Corresponding author: aad@utt.fr, lerondel@utt.fr

+ Current address: Laboratoire de Chimie des Matériaux Organiques, Thales Research and Technology, 1 avenue Augustin Fresnel, 91767 Palaiseau cedex, France

**I - Introduction**

Zinc oxide (ZnO) is a semiconducting material characterized by a direct, wide bandgap (~3.4 eV) and a relatively high exciton binding energy (60 meV). Together, these attributes offer the attractive perspective of low-threshold UV stimulated emission at room temperature (RT) and above **[1-4]**. Photoluminescence (PL) is a common non-destructive tool for characterizing ZnO thin films and nanostructures. In conventional RT PL studies, spectra exhibiting an excitonic near band edge (NBE) main emission peak and a relatively low "green signal" are commonly taken as the signature of high materials quality [**5**]. Such PL studies are primarily performed with normal-incidence collection, without regard to the angular dependence of the emission.

However, in the most commonly studied system of ZnO thin films on c-sapphire substrates, the refractive index of the ZnO layer at the excitonic peak emission wavelength (375 nm) is greater than that of the substrate, respectively 2.5 and 1.8 **[6, 7]**. Therefore waveguiding can be anticipated. Thus, in-plane emission would be expected to be subject to micro-cavity effects and Fabry-Perot-like interferences, which depend strongly on the wavelength ($\lambda$), film thickness and polarization. Study of the angular dependence of such ZnO emission is therefore, necessary in order to develop a proper and comprehensive understanding of the emission properties.

This paper focuses on measuring and modelling ZnO side-emission which is found to be dominated by leaky substrate modes. The novel analysis and simulation approach developed herein demonstrates that study of leaky modes can yield essential material properties, such as refractive index, film thickness and more importantly, the intrinsic emission profile (bulk emission).

**II – Experiment**

The ZnO layer studied in the paper was grown on a c-sapphire substrate using pulsed laser deposition (PLD), as described elsewhere **[8]**. Surface morphology was investigated using a Park Scientific Instruments M5 nanoprobe atomic force microscope (AFM) in tapping mode. The crystal structure was investigated by X-Ray diffraction (XRD) conducted in a Panalytical MRD-PRO system with Cu K$\alpha$ source and high resolution optics. Angle and Polarization-Resolved Photoluminescence (ARPL and PRPL) experiments were conducted on the sample. The sample was pumped at normal incidence by a continuous wave (CW) He-Cd laser emitting at 325 nm with a power density of less than 200 mW/cm$^2$. Luminescence was collected using a lens with a 0.13 NA focused onto a large core (400 μm diameter) optical fiber connected to a 50 cm focal length spectrometer. For ARPL, the collection aperture was primarily focused on the sample center but also, occasionally, on the sample edge. The collection configuration is subsequently indicated in the inset of each ARPL figure. As for PRPL, the collection aperture was always directed toward the sample edge. The collection optics were mounted on a goniometer in order to vary the collection angle ($\theta$).

**III – Theoretical background**

A. Emission Modelling

In order to simulate emission spectra, two distinct numerical models were employed in this paper. The first one consists of a simple opto-geometrical approach based on transmission calculations. The second aims at calculating the emission using the source terms (ST) model.

A.1. Transmission based model

The radiated field outside the thin film is here modeled as a summation of the direct transmission and the transmission after multiple reflections in the thin film (Fig.1-a). Such multi-beam summation gives the angle and wavelength dependences of the interference that determines the overall transmission. Considering a point source inside the film, the intensity transmitted at the substrate interface, for both transverse electric (TE) (i.e. electric field vector parallel to the film plane) and transverse magnetic (TM) (i.e. magnetic field vector parallel to the film plane) polarized light, is, therefore, expressed by [9]

$$T^{s,p}(\theta, \lambda) = \frac{n_{sub} \cos(\theta_{sub})}{n_{film} \cos(\theta_{film})} \left| \frac{t_2^{s,p}}{1 - \exp(2i\phi) r_1^{s,p} r_2^{s,p}} \right|^2 \qquad (1)$$

where

$$\phi = \frac{2\pi n_{film}}{\lambda_0} L \cos(\theta_{film}). \qquad (2)$$

Here, $r$ and $t$ represent Fresnel coefficients for reflection and transmission, while the 1 and 2 subscripts and the s and p superscripts indicate whether these coefficients correspond to film-air and film-substrate interface and to TE and TM polarized light, respectively. In addition, $n$ and $\theta$ refer to the refractive index and the propagation angle of emission (i.e. angle of incidence), while the *film* and *sub* subscripts indicated whether they are relative to the film or the substrate, respectively. Finally, L and $\lambda_0$ indicate the film thickness and the wavelength in vacuum (i.e. in air), respectively.

Eq. (1) gives the thin film transmittance for a specific angle and λ. However, PL measurements are collected through an objective lens, which is characterized by a numerical aperture (NA). Collected spectra, therefore, are the result of an accumulation of photons entering the lens over the solid angle defined by the NA. To account for this in the transmittance simulation, the λ-dependent transmittance was obtained by summing the transmissions over all angles whose sine was smaller than the NA ($\theta - \theta_{NA} \leq \theta \leq \theta + \theta_{NA}$ with $\sin\theta_{NA}$=NA)

$$T^{s,p}(\lambda) = \sum_{\theta-\theta_{NA}}^{\theta+\theta_{NA}} T^{s,p}(\theta,\lambda) \tag{3}$$

A.2. Source Terms model

The second model, the ST method **[10]**, is known for simulating the emission from arbitrary planar structures. Compared to other approaches, the method has the advantage of being simple and convenient for calculating the emission characteristics of dipoles embedded in planar media. The ST method unifies dipole emission and standard matrix methods by adding explicit ST. The model considers 3 electrical dipole configurations, 1 vertical (*v*, along z axis) and 2 horizontal (*h*, along x and y axis). Dipoles are distributed along a sheet (plane ⊥ to z axis) forming the source plane (Fig. 1-b). For each dipole configuration, a source term and a polarization are attributed (see Table 1). *A* presents the source terms and the superscripts denote the propagation direction of the emission in the z direction : > for the propagation towards the substrate and < for the propagation towards the air.

Table 1. Source terms for horizontal and vertical dipoles.

| Dipole | Mode | |
|---|---|---|
| | TE | TM |
| Horizontal | $A^{>,<} = \pm\sqrt{\dfrac{3}{16\pi}}$ | $A^{>,<} = \pm\sqrt{\dfrac{3}{16\pi}}\dfrac{k_{z,film}}{k_{film}}$ |
| Vertical | $A^{>,<} = 0$ | $A^{>,<} = \pm\sqrt{\dfrac{3}{8\pi}}\dfrac{k_{\parallel}}{k_{film}}$ |

Where $k_{film}$ is the wavevector inside the film and $k_{z,film}$ and $k_{//}$ are its projection along the z axis and the xy plane, respectively.

Calculating the external field for a multilayered medium then comes down to solving the following system of equations

$$\begin{pmatrix} E_2^< \\ E_2^> \end{pmatrix} - \begin{pmatrix} E_1^< \\ E_1^> \end{pmatrix} = \begin{pmatrix} A^< \\ A^> \end{pmatrix}, \tag{4}$$

$$\begin{bmatrix} a_{11} & a_{12} \\ a_{21} & a_{22} \end{bmatrix} \begin{pmatrix} 0 \\ E_{air} \end{pmatrix} = \begin{pmatrix} E_1^< \\ E_1^> \end{pmatrix}, \tag{5}$$

$$\begin{bmatrix} b_{11} & b_{12} \\ b_{21} & b_{22} \end{bmatrix} \begin{pmatrix} E_{sub} \\ 0 \end{pmatrix} = \begin{pmatrix} E_2^< \\ E_2^> \end{pmatrix}. \tag{6}$$

Where $E_1$ and $E_2$ are the magnitudes of the electrical field on the right side and the left side of the source plane, while $E_{air}$ and $E_{sub}$ are the magnitudes of the electrical field radiated in air and sapphire, respectively. (a) and (b) are 2 x 2 (transfer) matrices depicting the propagation towards the source plane from the air and the substrate, respectively (Fig. 1-b).

Eq. (4) expresses an electrical field discontinuity at the source plane induced by the existence of electrical dipoles (i.e. electrical charge). On the other hand, Eq. (5) and Eq. (6) describe the relation linking the externally-radiated and internally-emitted fields.

The extracted power in the substrate ($P_{extracted}$) is then given by:

$$P_{extracted} \propto \int_{\theta-\theta_{NA}}^{\theta+\theta_{NA}} \Pi(\theta)\sin(\theta)d\theta \tag{7}$$

with

$$\Pi(\theta) = |E_{sub}|^2 \frac{n_{sub}k_{z,sub}^2}{n_{film}k_{z,film}^2} \tag{8}$$

where $k_{z,film}$ and $k_{z,sub}$ represent the z-wavevector in the thin film and the substrate, respectively.

As in the first method, the integral giving $P_{extracted}$ is defined over all the angles entering the lens ($\theta-\theta_{NA} \leq \theta \leq \theta+\theta_{NA}$).

Furthermore, the ST method considers the emission of the whole film, as opposed to the transmission calculation, which considers a point-like source. In the case of a finite volume, source planes can be distributed along the layer thickness. The total radiated power is then obtained by summing the emitted power of the discrete subsources.

To summarize the modelling section, the transmission calculation only considers a classical approach to calculate the emission. The transmission simulation has the advantage of being simpler and faster than the ST method but it only simulates the cavity response for a non-emitting optical thin film. ST calculations, on the other hand, take into account the contribution of dipolar moments in addition to opto-geometrical effects. The ST model also considers the emission from the whole ZnO volume, while the transmission model only considers a point-like light source. The ST method is, therefore, the most appropriate method and, as shown later, it also allows retrieval of the intrinsic emission of an active thin film.

B. Leaky modes

In asymmetric slabs (i.e. $n_{air} < n_{sub} < n_{film}$) such as ZnO on sapphire, emitted light can couple into leaky substrate modes. Depending on the angle of incidence, reflectivity at each thin film interface varies. At grazing substrate angles (e.g. $\theta_{sub} \approx 90°$, Fig. 1-a), total internal reflection occurs at the film/air interface (e.g. reflectance, $r_1^{s,p} = 1$, Fig. 1-a); while reflection at the film/substrate interface is strong but only partial (e.g. reflectance, $r_2^{s,p} \approx 1$, Fig. 1-a). The thin film emission that is transmitted, or in other words 'leaks', into the substrate then propagates in it near the film/substrate interface and escapes (and therefore is collected) at the substrate edge. Such quasi-guided slab modes are called 'leaky modes' [**11**]. Because of their quasi-waveguided nature, leaky modes are subjected to strong Fabry-Perot interference which dominates the side emission spectra. At grazing angles, the thin film presents a microcavity behavior as its interfaces are comparable to two mirrors with $r_1^{s,p} = 1$ and $r_2^{s,p} \approx 1$. On the other hand, for smaller angles, the microcavity behavior eventually vanishes as the interface reflectivity is gradually reduced. Consequently, Fabry-Perot interferences [e.g. Eq. (1)] are more prominent when collecting from the substrate side. Calculating the cavity response (i.e. transmittance) is, therefore, a prerequisite for understanding the side emission and analyzing PL spectra.

**IV – Results and discussion**

Optical interferences in emission are likely to be observed in epitaxial thin films due to the interface quality. XRD 2θ/ω scans revealed the epitaxy of ZnO on the sapphire substrate. The main (0002) peak exhibited strong Pondellusong fringes, which were consistent with the film having a relatively smooth surface over the scale of the X-Ray spot (few mm$^2$) and gave an estimate of film thickness at 160 nm. ω rocking curves for the (0002) peak gave a linewidth of < 100 arcsec, indicating a highly oriented film with relatively low crystallographic dispersion about the crystallographic c-axis. AFM confirmed the ZnO layer to have a relatively low root mean square roughness of < 2 nm. Based on the characteristics, Fabry-Perot fringes are expected in such a film.

Fig. 2 presents the ARPL intensity plot as a function of collection angle ($\theta$), together with a sketch of the experimental setup. The presence of an emission at $\theta$ greater than 90° is clearly revealed. One may also notice the minimum in emission intensity is at 90°. This can be explained by the fact that the guided mode within the thin film is almost completely attenuated. Guided modes are strongly affected by the absorption losses unless gain is generated using higher pumping intensities. Gain has been already observed in epitaxial grown ZnO thin films [12] and more especially for the PLD grown sample similar to the one studied here with a typical power density threshold, $D_{th}$ > 40kW/cm$^2$ [**6, 12, 13**]. Working with a CW laser implies that the edge emission ($\theta > 90°$) cannot be attributed to guided modes but is most probably dominated by leaky modes for which losses are less important. Before going in more details of the origin of the emission, edge emission is further investigated.

Fig. 3 shows a contour mapping of the normalized PRPL spectra as a function of $\theta$ and λ for both the TE and TM polarizations. The plots show sharp and well-separated TE and TM luminescence peaks ('fine peaks') observed for a narrow angular domain between 90° and 105°; while, for angles bigger than 105°, the TE and TM luminescence profiles are broader, overlap more and resemble the surface emitted spectrum (i.e. Fig. 4-a). The sharpest luminescence peaks are observed for $\theta$ near 90°. This behaviour seems similar to the one observed for similar luminescent thin films reported in the literature **[9, 14, 15]**. A clear signature of a leaky mode emission and a microcavity behaviour that is more prominent at grazing angles is also clearly observed. This is due to the high reflectivity at the ZnO interfaces (cf § III-B).

Fig. 4-a and Fig. 4-b show the ARPL spectra recorded for 2 different collection angles ($\theta = 15°$ and 95°). Surface emission (Fig. 4-a, 15°) gave a single main peak, which was attributed to the excitonic NBE emission. However, it can be seen that a substantially different spectrum is recorded (Fig. 4-b, 95°) from the back-side of the sample (i.e. $\theta > 90°$). The spectrum then consists of a doublet with maxima at 381 and 388 nm. Previous ARPL spectra (Fig. 4-a and Fig. 4-b) were collected by directing the collection towards the centre of the sample, as shown in the figure inset. However, when the collection is redirected (at the same $\theta$) towards the edge of the sample there is a significant reduction in the width of the emission peaks (e.g. Fig. 4-b and Fig. 4-c). This can be attributed to improved spatial filtering of the leaky mode emission, which usually propagates close to the film/substrate interface. The full radiated emission is then efficiently separated from the leaky mode emission. PRPL was then used in order to confirm the origin of the emission. PRPL showed that the doublet peaks actually have different polarizations, as can be seen in Fig. 4-d. The first peak wavelength maximum ($\lambda_{MAX} \sim 381$ nm) was mainly TM polarized, while the second peak $\lambda_{MAX}$ (~ 389 nm) was mainly TE polarized. It is also observed that the TE peak is more intense by a factor of about 1.9. These peaks are linearly polarised and have a much lower Full Width Half Maxima (FWHM), of 4.4 and 4.1 nm, respectively, than the peaks measured

far from the edge at the middle of the sample (cf. Fig. 4-b). The polarized fine peaks are again consistent with microcavity behaviour in the ZnO thin film.

The thin-film cavity response was then simulated using the opto-geometrical model (cf. § III-A.1) in order to confirm the origin of the fine structures observed in emission. Fig. 5 shows the theoretical spectra calculated for a 95° $\theta$ with a 15° aperture.

For refractive indices (*n*) of 2.4 for ZnO and 1.9 for sapphire, TE and TM calculations indeed showed polarized fine peaks (Fig. 5-a). However, the theoretically predicted spectral positions and relative intensities of the two resonances do not match the experimental results. A much better agreement was found for the spectral positions and FWHM when the dispersion and the anisotropy of the ZnO complex *n* were taken into account **[16]**, (Fig. 5-c). While the agreement in terms of peak positions already allows for the retrieval of the opto-geometrical parameters the agreement can be further improved. As shown in Fig. 5, the theoretical TE and TM amplitude still did not correlate with the experimental data. As shown later, this could be due to the intrinsic ZnO emission and/or dipole anisotropy, which the model does not take into account. The transmission model simply calculates the transmission factor for the thin ZnO layer. Nonetheless, the fact that the model accurately predicts the peak features indicates that emission emanating from the edge of the sample is more influenced by the properties of the planar waveguide cavity than by the intrinsic properties of the material (bulk emission). The good agreement obtained between the experimental spectra and the calculations is consistent with leaky modes being the origin of the side emission. Interestingly, although leaky modes are evanescent in air, they can still be observed from the surface at normal collection as shown by the peak at 388 nm (indicated by the arrow in Fig. 4-a). This is probably due to light scattering induced by the surface roughness, which can partially convert the modes in radiating waves towards the air. This feature further confirms the nature of the observed side emission.

At this stage, it is worth noting that in this paper, only parameters taken from the literature **[16]** (i.e. non adjustable) were used. This in turn validates the reported method.

The thin-film transmission based model presented above simulates a cavity response which is coherent with the fine PL peaks originating from a cavity-interference effect. However, the model does not fully describe the PL spectra, since it is for passive optical materials (non emitting material). In order to take emission into account, extended calculations were made using the ST method (cf. § III-A.2) which was recently adapted to model waveguiding in ZnO thin films similar to these probed in this work **[17]**.

Fig. 6-a presents the luminescence spectra calculated using the same simulation parameters as for Fig. 5-c. The spectra shown in Fig. 6-a are again consistent with an opto-geometrical origin for the polarized fine peaks. The calculated peaks both agree with the experimental data. However, the calculated TE spectrum shows a better agreement with the experimental data than the calculated TM spectrum. This can be explained by the wavelength dependence of the bulk ZnO emission efficiency that the ST model has not accounted for. As the bulk ZnO emission $\lambda_{MAX}$ is normally ~375nm, the TM mode ($\lambda_{MAX}$ ~380 nm) is more likely to be influenced than the TE mode ($\lambda_{MAX}$ ~390 nm). In a simplified approach, the thin film emission spectrum can be considered to be the convolution of the bulk ZnO emission and the ST spectra. The emission can, therefore, be written as $E_{TF}(\lambda) = E_{bulk}(\lambda).ST(\lambda)$, where $E_{TF}$, $E_{bulk}$ and $ST$ are, respectively, the thin film emission, the bulk emission and the ST calculation for the TE and TM polarization **[18]**. The results shown in Fig. 6-a were obtained without considering the λ dependence of the bulk material emission (i.e. an $E_{bulk}(\lambda) = 1$ for all λ).

In an inverse approach, by comparing the experimental and simulated spectra, the ST method can be used to retrieve the bulk emission of the material i.e. a λ-dependent weighting coefficient. This coefficient is obtained by calculating the ratio between the experimental and the simulated spectra. Fig. 6-b presents the results obtained for the TE and TM polarizations. For shorter λ (higher energies) the cavity effect is negligible, due to strong absorption in ZnO, and, therefore, no information can be gained on the ZnO luminescence. Hence, for shorter λ, Fig. 6-b shows only noise. For higher λ, TE and TM peaks appear. Most remarkably, for the TE mode, an emission peak, not clearly visible in the experimental spectrum, is revealed. This peak is at a $\lambda_{MAX}$ of ~376 nm (~3.3 eV), which corresponds to excitonic NBE emission. This emission was already experimentally observed, as indicated by the arrow on the left in Fig. 4-d. The TE efficiency is almost 3 times higher than for the TM. This is in good agreement with the TE/TM emission anisotropy already reported. It is widely reported that ZnO dipoles are preferentially TE polarized for thin film grown on c-sapphire [**19, 20**]. The experimentally observed TE:TM anisotropy is 2:1 (Fig. 4-d). This anisotropy is not that of the ZnO bulk emission, but that of the cavity modes, as shown in Fig. 6. The TM mode, which exists at almost the same λ as the bulk emission, helps in enhancing the TM-polarized ZnO excitonic emission. This is not the case for the TE polarization, as TE cavity resonance exists at higher λ (~388 nm).

The strong resemblance of the emission spectra in Fig. 6-b (TE+TM) to those obtained at small *θ*, Fig. 4-a, validates the ST method. In fact, at small *θ* (i.e. small angle of incidence), both the ZnO-air and ZnO-sapphire interfaces exhibit low *r* and, therefore, the ZnO film forms a poor quality cavity. As a consequence, the cavity effect on the ZnO intrinsic emission is weak but, nevertheless, always exists. This can lead to additional peaks, as evidenced very recently for relatively thick (750 nm) Al doped ZnO films deposited on a relatively low *n* substrate (quartz) **[21]**. The spectra obtained in Fig. 6-b can thus reveal the bulk emission through leaky mode analysis. It is important to note again that the bulk emission is obtained without any adjustable parameters.

**IV - Conclusion**

In this work, the dependence of PL spectra on $\theta$ for ZnO epitaxial thin films grown on c-sapphire substrates by PLD was investigated using ARPL and PRPL. Simulation succeeded in describing how the emission spectra collected from the edge of the sample (at angles higher than 90°) were strongly affected by the excitation of leaky modes radiating into the substrate at grazing angles. These leaky modes were modeled as Fabry-Perot-like interferences coming from multiple total and partial reflections between the ZnO/air and ZnO/sapphire interfaces. It was also shown that such leaky modes can trap the luminescence from the ZnO layer and lead to λ-selective constructive interference, which produces a narrowing of the initial luminescence spectrum. Moreover, calculations show that $n$ dispersion and anisotropy have to be considered in order to correctly account for the observed experimental spectra. Thus it was deduced that leaky mode analysis is highly sensitive to the opto-geometrical parameters of the thin films, which means that the technique is a very precise method for the estimation of these parameters. In addition it was shown, using simulation, that the measured spectra can be deconvoluted and that the ZnO bulk emission can be deduced. While optical cavity effects, especially leaky modes, are more clearly observed at higher $\theta$, normal incidence emission can also be affected (depending on the roughness of the thin film). As a direct consequence, it is concluded that fine spectroscopic analysis of the emission from thin films deposited on low $n$ materials should take into account optical cavity effects as a general rule.

This work was partially supported by the ANR project ULTRAFLU and the Champagne-Ardenne regional council project MATISSE. Two of the authors, L. D. and R. A. would also like to thank the Champagne-Ardenne regional council and the ERDF (European Regional Development Fund) program for post-doctoral and doctoral support, respectively.

**FIGURE CAPTIONS**

Fig. 1. (a) Sketch of the ray path for direct transmission and after reflection at both thin film interfaces. (b) Sketch of a thin film structure with an internal source plane and matrices describing the propagation of the electric fields outside the source.

Fig. 2. Sketch of the experimental setup with the collection configuration and the chosen origin for the $\theta$ together with the evolution of the luminescence intensity as a function of $\theta$. The sample was excited at normal incidence.

Fig. 3. Normalized angular luminescence mapping of both TE (a) and TM (b) modes. Arrows indicate the spectral narrowing and shift between leaky modes and substrate emission.

Fig. 4. (a) Normalized luminescence spectra recorded at 15°. The inset shows a sketch setup with the collection aperture focused on the centre of the sample. (b) Normalized luminescence spectra recorded at 95° with a collection spot coincident with the excitation spot. The inset shows a sketch of the collection aperture focused on the backside of the sample. (c) Normalized luminescence spectra recorded at 95° with a collection spot located at the side of the sample. The inset shows a sketch of the collection aperture focused on the backside edge of the sample. (d) PRPL spectrum for the film luminescence at $\theta=95°$.

Fig. 5. Transmission based model simulations for TE (red line) and TM (black line) polarized light. (a) Simulation results for fixed ZnO and sapphire $n$ of 2.4 and 1.9, respectively. (b) Simulation results taking into account the ZnO $n$ with no anisotropy. (c) Simulation results

taking into account the ZnO $n$ dispersion and anisotropy. Simulation results are compared with experimental PL (squares) in Fig. 4-d.

Fig. 6. (a) Simulated PL spectra by ST method for TE (red line) and TM (black line) polarized light. The simulated spectra are compared with the experimental PL (squares) in Fig. 4-d. (b) Ratio of the experimental and simulated PL i.e. the bulk emission.

**FIGURES**

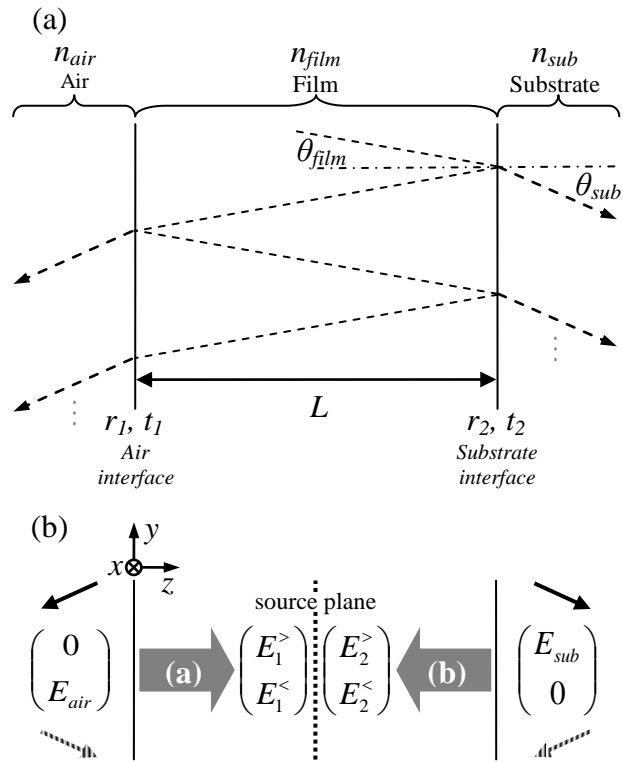

Fig. 1

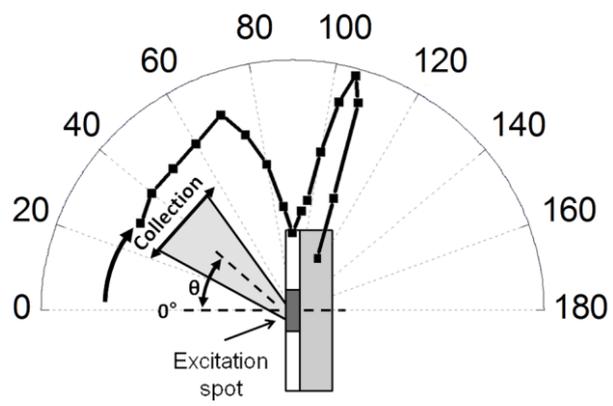

Fig. 2

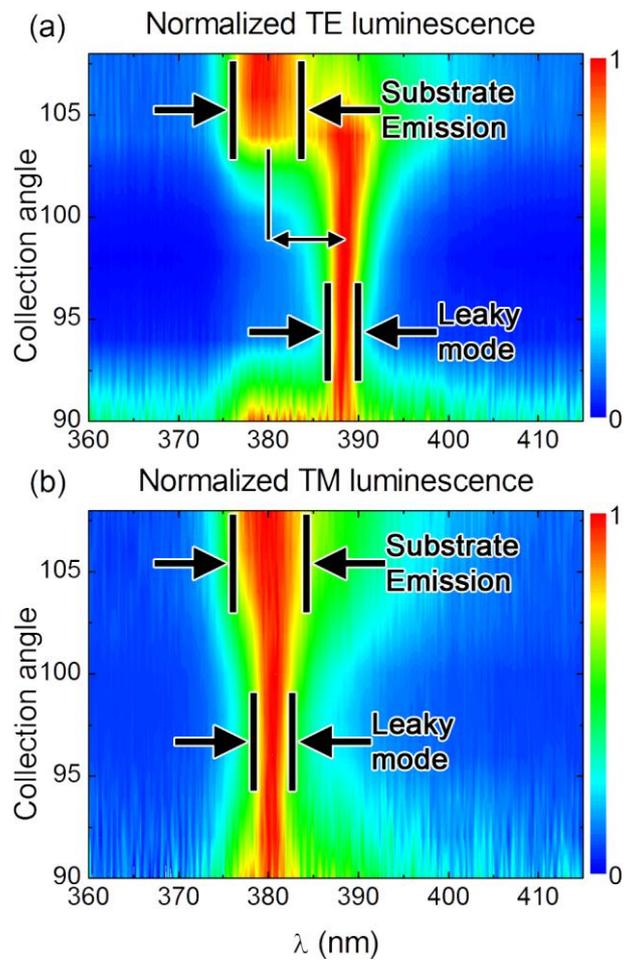

Fig. 3

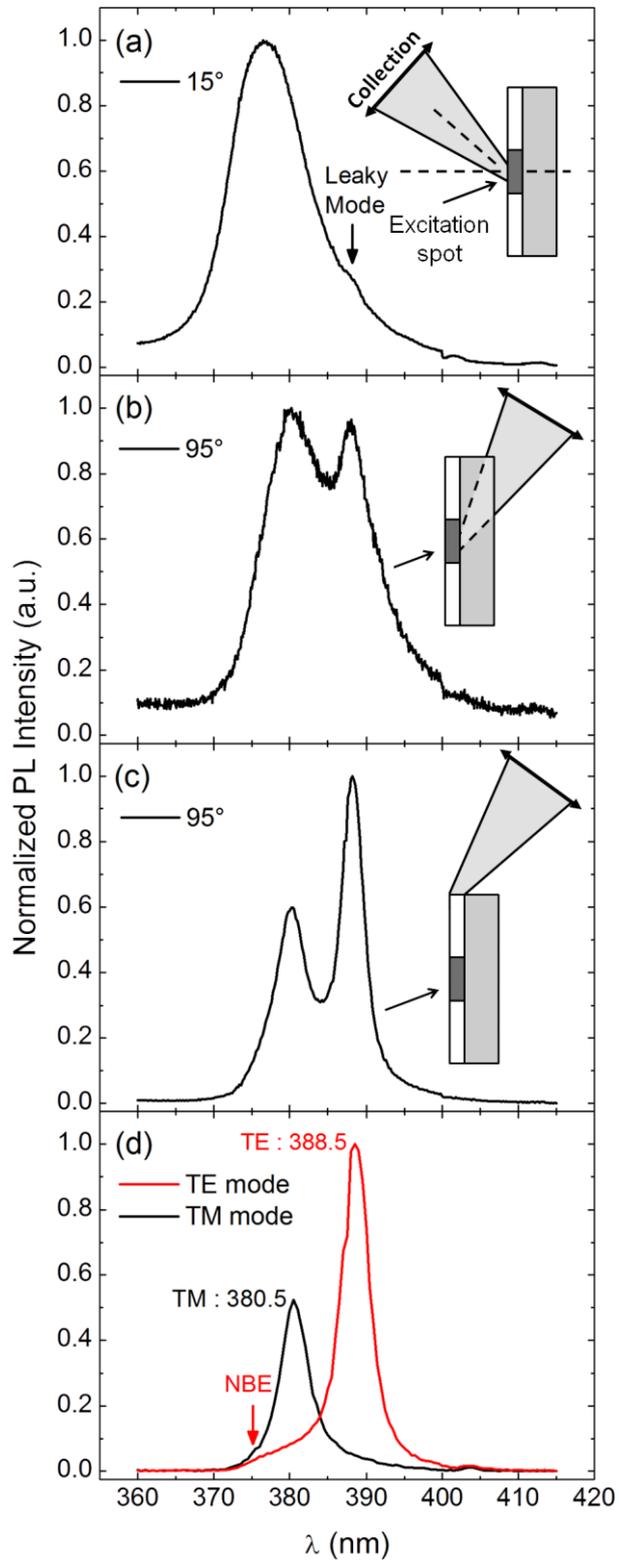

Fig. 4

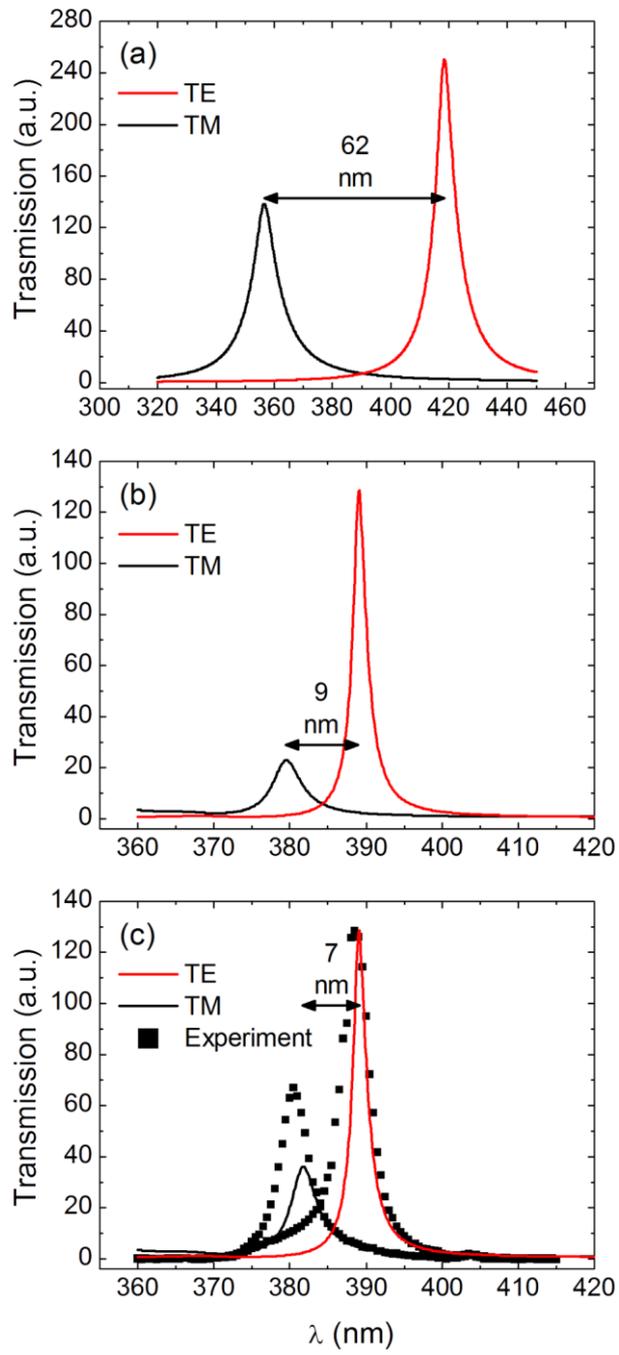

Fig. 5

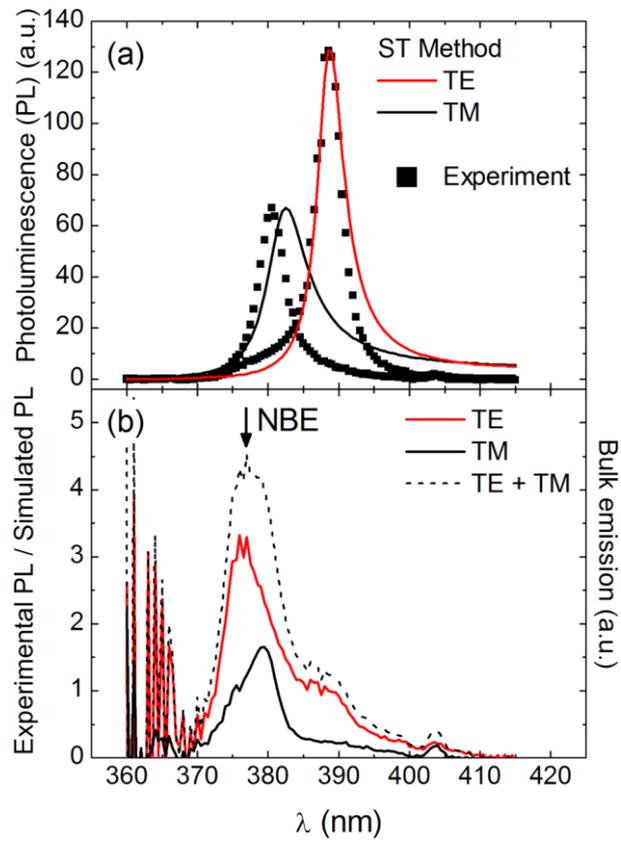

Fig. 6